\documentclass[a4paper,11pt]{article}
\usepackage{amsmath}
\usepackage{amsfonts}
\usepackage{amssymb}
\usepackage{bm}
\usepackage[utf8x]{inputenc}
\usepackage{ae}
\usepackage[T1]{fontenc}
\usepackage[english]{babel}
\usepackage{graphicx}

\usepackage{makecell}
\usepackage{placeins}
\usepackage{tabularx}
\usepackage{tabulary}
\usepackage{setspace}
\usepackage{tablefootnote}
\usepackage[flushmargin]{footmisc}
\usepackage[comma]{natbib}
\usepackage{array}
\usepackage{float}
\usepackage{caption}
\usepackage{tikz}
\usetikzlibrary{shapes.geometric}
\usepackage[a4paper,left=2.5cm,right=2.5cm,top=3cm,bottom=3cm]{geometry}
\usepackage[colorinlistoftodos]{todonotes}
\usepackage{pdfpages}
\usepackage{longtable}
\usepackage{adjustbox}
\usepackage{booktabs}
\usepackage{multirow}
\usepackage{lscape}
\usepackage[colorlinks=true, allcolors=blue]{hyperref}

\usepackage[flushleft]{threeparttable}
\usepackage{inputenc}



\begin{document} \doublespacing \pagestyle{plain}
	
	\def\ci{\perp\!\!\!\perp}
	\begin{center}
		
		{\LARGE Predicting Children's Travel Modes for School Journeys in Switzerland: A Machine Learning Approach Using National Census Data}
		
		{\large \vspace{0.8cm}}
		
		{\large Hannes Wallimann and Noah Balthasar }\medskip

		{\small {University of Applied Sciences and Arts Lucerne, Institute of Tourism and Mobility} \bigskip }
		
		{\large \vspace{0.8cm}}
		
		{\large Version: April 2025}\medskip
		
	\end{center}
	
	\smallskip

	\noindent \textbf{Abstract:} {Children’s travel behavior plays a critical role in shaping long-term mobility habits and public health outcomes. Despite growing global interest, little is known about the factors influencing travel mode choice of children for school journeys in Switzerland. This study addresses this gap by applying a random forest classifier---a machine learning algorithm---to data from the Swiss Mobility and Transport Microcensus, in order to identify key predictors of children’s travel mode choice for school journeys. Distance consistently emerges as the most important predictor across all models, for instance when distinguishing between active vs. non-active travel or car vs. non-car usage. The models show relatively high performance, with overall classification accuracy of 87.27\% (active vs. non-active) and 78.97\% (car vs. non-car), respectively. The study offers empirically grounded insights that can support school mobility policies and demonstrates the potential of machine learning in uncovering behavioral patterns in complex transport datasets.}
	
	{\small \smallskip }
	{\small \smallskip }
	{\small \smallskip }
	
	
	{\small \smallskip }
	{\small \smallskip }
	{\small \smallskip }
	
	{\small \noindent \textbf{Acknowledgments:} We are grateful to the "IDN Gesundheit" at the Lucerne University of Applied Sciences and Arts for their financial support. We also thank Timo Ohnmacht and the other team members of the Competence Centre for Mobility at the Lucerne University of Applied Sciences and Arts for their valuable input and insightful discussions. The authors acknowledge the use of ChatGPT (OpenAI) and Grammarly to support text refinement, data processing, and table creation. All analyses, interpretations, and final writing decisions are solely the responsibility of the authors.}
	
	\bigskip
	\bigskip
	\bigskip
	\bigskip
	
	{\small {\scriptsize 
			\begin{spacing}{1.5}\noindent  
				\textbf{Addresses for correspondence:} Hannes Wallimann, University of Applied Sciences and Arts Lucerne, Rösslimatte 48, 6002 Lucerne, \href{mailto:hannes.wallimann@hslu.ch}{hannes.wallimann@hslu.ch}; Noah Balthasar, \href{mailto:noah.balthasar@hslu.ch}{noah.balthasar@hslu.ch}.
			\end{spacing}
			
		}\thispagestyle{empty}\pagebreak  }

	{\small \renewcommand{\thefootnote}{\arabic{footnote}} %
		\setcounter{footnote}{0}  \pagebreak \setcounter{footnote}{0} \pagebreak %
		\setcounter{page}{1} }
	
	\section{Introduction}\label{introduction}
	
	Motorized traffic restricts children’s commuting patterns and outdoor activities, leading to negative health consequences by reducing physical activity and overall well-being. Moreover, transporting children to school with private cars often causes drop-off areas to become highly congested and reduces pedestrian visibility, making the environment less safe for active travel. In contrast, children’s active and independent mobility between home and school is an opportunity for children to achieve regular physical activity and improve well-being, create safer and quieter environments, and reduce traffic congestion. However, the proportion of children walking to school declined around the globe in many countries \citep[see, among others, ][]{mcmillan2007relative,curtis2015built,cadima2024walkability}.

	The travel mode choice of children---particularly with regard to parental driving between home and school---is influenced by a variety of factors. School travel patterns are shaped by a complex interplay of individual, familial, and environmental conditions. While numerous studies have been conducted worldwide \citep[see, e.g.,][]{mcmillan2007relative,panter2008environmental,fyhri2011children}, analyses focusing on Swiss children remain relatively rare, despite it being an interesting case due to diverse language regions and regional variation in mobility cultures. One exception is the study by \citet{grize2010trend}, which examines the time trends in active commuting among Swiss children and the factors associated with changes in walking and biking between 1994 and 2005. Another is the study by \citet{sauter2019}, which provides a descriptive account of children's mobility in Switzerland based on data from 1994 to 2015. However, despite various policies implemented by municipalities and schools in Switzerland---mainly aimed at reducing the use of private cars \citep[see, e.g., ][]{balthasar2024fare}---there remains a research gap concerning the determinants that influence children's travel mode choice for school journeys, especially active travel modes, and parental cars. 
	
	To address this gap, we apply machine learning methods to a large and comprehensive dataset of about 3,000 observations from the Swiss national travel census. Using a random forest classifier, we identify the most important predictors of children's travel mode choice for school journeys and highlight how different factors contribute to the likelihood of being driven to school versus choosing other modes. Furthermore, we compare the predictive performance and variable importance when distinguishing between individual travel modes and when grouping modes—for example, by comparing active versus non-active travel or car versus no-car. Our paper not only offers new insights into the determinants of school travel in Switzerland but also demonstrates the added value of machine learning in capturing complex patterns in mobility behavior.

	Distance to school emerges as the most influential variable across all prediction models for predicting children's travel modes for school journeys. Other relevant factors include the child’s age, the number of bicycles in the household, whether the individual lives in the French-speaking part of Switzerland, household size, and whether the school is located in the same municipality.
	The results indicate that when predicting active travel (i.e., biking or walking), the overall classification accuracy amounts to 87.27\%, with a 95\% prediction interval of [86.82\%; 87.71\%]. The accuracy slightly decreases when predicting motorized traffic to school (78.97\%), with a 95\% prediction interval of [78.28\%; 79,67\%]. However, we still achieve a reasonable classification of car usage of 70.80\% with a prediction interval of [69.94\%; 71.67\%] when excluding distance as a predictor.
	This study provides important insights for informing policy measures aimed at promoting active travel and reducing motorized traffic to schools.

	The paper proceeds as follows. In Section \ref{literature}, we present the relevant literature on children's travel mode choices for school journeys. In Section \ref{Methods}, we present the random forest classifier we use to predict the travel mode choice of children for school journeys. Section \ref{Desc} presents the census data from Switzerland and descriptive statistics. In Section \ref{Results}, we show the results. Finally, Section \ref{Discussion} discusses and concludes.

\section{Literature Review}\label{literature}

Children’s travel mode choice for school journeys---particularly parental driving---is influenced by a variety of factors. School travel patterns are shaped by a complex interplay of individual, familial, and environmental influences, including distance, infrastructure, safety perceptions, and social norms \citep[see, e.g.,][]{mcmillan2005urban,panter2008environmental}. In the following, we present these factors and group them into five categories: distance to school and travel time, built environment, socio-demographic characteristics, parental and household characteristics, and perceived safety and risk factors.

\bigskip

\textit{Distance to school and travel time}

The most frequently---and often most significantly---mentioned factors influencing walking to school are distance and travel time. \citet{mcdonald2008children}, analyzing data from the U.S. National Household Travel Survey (NHTS), shows that as travel time increases, the likelihood of walking to school decreases. This finding has been replicated in a wide range of studies. Examples include \citet{waygood2015walking}, who analyze the Scottish Household Survey, and \citet{fyhri2011children}, who compare the development of children’s mobility in Denmark, Finland, Great Britain, and Norway. In a case study of Ireland, \citet{kelly2014sustainable} identify 2 km as a guiding threshold: above this distance, primary school children rarely walk to school.

\bigskip

\textit{Built environment}

Other important factors influencing active travel for school journeys relate to the built environment and infrastructure. In areas with better pedestrian and cycling infrastructure, active school travel is more common \citep{cadima2024walkability}. On the one hand, safer routes with appropriate crossings and less busy roads are positively associated with active travel \citep{panter2008environmental}. On the other hand, the absence of sidewalks, pedestrian crossings, and road safety measures discourages walking and cycling \citep{timperio2006personal}. Moreover, dense and pedestrian-friendly residential areas encourage mobility; therefore, children in urban areas with good access to schools are more likely to walk or cycle \citep{curtis2015built}. However, even in compact urban settings, parents often escort their children due to perceived dangers, which reduces opportunities for independent travel \citep{katsavounidou2024active}.

\bigskip

\textit{Socio-demographic characteristics}

Moreover, various socio-demographic characteristics influence children’s mode choice for school journeys. For instance, older children are more likely to travel actively than younger ones \citep{fyhri2009children}. In addition, girls are more frequently escorted by parents in cars, while boys are (slightly) more likely to engage in active travel and enjoy greater mobility freedom \citep{mcmillan2006johnny,stewart2011findings}. Another factor that may influence travel mode choice is ethnicity---although travel distance may also vary by ethnic background \citep{easton2015children}. Moreover, the study of \citet{grize2010trend} that children from French-speaking regions in Switzerland were more likely to be driven to school compared to their German-speaking counterparts. 

\bigskip

\textit{Parental and household characteristics}

In addition to the socio-demographic characteristics of the child, parental and household characteristics also influence the choice of travel mode. This includes parental socio-economic or employment status and education \citep{cadima2024walkability,stewart2011findings}. Household size might be relevant, as siblings can facilitate active travel because they provide someone to accompany the younger child when walking to school \citep{ahlport2008barriers}. 
Furthermore, studies show that household car ownership reduces the probability of active school travel \citep{mcdonald2005children,mcdonald2011us,mcmillan2003walking}.

\bigskip

\textit{Perceived safety and risk factors}

Further, congested school drop-off zones create safety risks for children walking or cycling \citep{mammen2012understanding}. Parents often cite traffic, unsafe crossings, and stranger danger as key concerns \citep{cadima2024walkability,westman2017drives}. These fears foster car dependence and limit children's independence \citep{cadima2024beyond}. Driving parents perceive the environment as more dangerous than those who support active travel \citep{johansson2006environment, hume2009children}. Beyond infrastructure, attitudes and affective motives strongly shape transport choices \citep{mcmillan2007relative, steg2005car}.

\bigskip

Finally, and in addition to these five categories, some studies also show that weather might be a predictor of active school travel \citep{chillon2014cross}. 

\section{Methods}\label{Methods}

In the present paper, we apply machine learning to predict children's travel mode choice for school journeys using a random forest classifier. Applying machine learning algorithms entails using a set of predictor variables ($X$) to predict a (binary) target variable ($Y$), in this case, the travel mode of children for school journeys. First, machine-learning approaches require splitting the data randomly into training and test sets, i.e., containing 75\% and 25\% of the observations, respectively. We then train predictive models using the training data, where both the outcome and predictors are known. The trained models are subsequently used to predict the outcome for the test data. The model's performance is estimated by comparing the predicted and true values of the target variable. To ensure robustness, we repeat the training and testing procedure 100 times by randomly splitting the data into test and training sets. This iterative evaluation allows us to calculate 95\%-prediction intervals, which provide insights into the model’s performance variability. Note that for every repetition, the data set in which the main travel modes for school journeys are balanced, e.g., with 50\% active (i.e., walking and cycling), and 50\% non-active (i.e., car and public transport) travel modes for school journeys. Balancing the data set each time before splitting the sample in training and test data enables the machine learning algorithm to build models which are able to predict all travel mode choice classes equally well. 

We implement the random forest as our machine learning algorithm (see \citet{breiman2001random} or \citet{imhof2023assessing} for an example in the transportation literature). In our case, this tree-based method builds multiple classifion trees by drawing random subsamples of the training set. The final classification decision is determined using a majority voting rule across all trees. Each individual classification tree predicts the travel mode by sequentially applying decision rules on predictor variables. The tree starts at a root node, applies a threshold for the most important variable, and moves down the corresponding branches based on the observation's feature values. This process continues until a terminal node (leaf) is reached, which provides the predicted travel mode for school journeys. However, individual decision trees often face a bias-variance trade-off: deeper trees reduce bias but increase variance, leading to potential overfitting. The random forest mitigates this trade-off by aggregating multiple trees and selecting a random subset of predictors at each node to reduce correlations among trees. Furthermore, random forests do not require fine-tuning of penalty terms \citep{athey2019machine}, making them relatively easy to implement. We use the 	\textit{randomForest} package in \textsf{R} by \citet{breiman2018randomforest} for our implementation.

An important advantage of the random forest classifier is its ability to identify the most important predictor variables. We compute variable importance scores, which measure the extent to which a predictor improves classification accuracy. These scores are based on the difference in average prediction error with and without the given predictor. Additionally, we analyze the relative importance of predictors using Mean Decrease Gini (MDG) scores. It is essential to note that this analysis does not establish a causal relationship between predictor variables and the actual travel mode choice for school journeys but rather evaluates the predictor’s effectiveness in forecasting mode choice. Machine learning prioritizes predictive accuracy, and correlations among predictors can influence the relative importance scores and partial dependence plots. Consequently, when interpreting results, we assume that predictors are independent.

\section{Data and descriptive analysis}\label{Desc}

\subsection{Mobility and Transport Microcensus}

In Switzerland, the Federal Statistical Office (FSO) and the Swiss Federal Office for Spatial Development (ARE) carry out the Mobility and Transport Microcensus (MTMC), a one-day computer-assisted telephone interviews (CATI) diary survey representative of the Swiss population \citep{FSO2017}. The aim of the MTMC is to collect information on the mobility behavior of the Swiss population. The survey is usually carried out every five years, with a postponement from 2020 to 2021 due to the COVID-19 pandemic. On behalf of the FSO and the ARE, a social research institute carries out CATI for 25 min on average. The sample represents the Swiss resident population in terms of selected socio-demographics and spatial distribution. The data collected consists of information on respondents’ daily mobility. The participants were asked to report all routes traveled on a specific reference date. Moreover, questions were asked regarding the availability of mobility tools---including vehicles and public transport season tickets---, socio-economic information on households and individuals, and occasional journeys (such as holidays) \citep[for more information on the MTMC, see also][]{weis2021surveying,ohnmacht2014route}. We use the most recent sample of 2021 that contains 55,000 individuals aged six years and older. Despite answering the questions voluntarily, the participation rate in 2021 amounted to 46\% \citep{FSO2017}. 

Overall, the 2021 sample contains 3,782 children younger than 12 who visit kindergarten or primary school (or a school that fulfills a similar role).

\subsection{Descriptive analysis}

For our analysis, we construct the dataset as follows: We consider outbound journeys\footnote{Note that we only consider outbound journeys (i.e., from home to school) to reduce avoid model overfitting. Moreover, the dataset of children's journeys does not distinguish between escorted and independent school journeys and does not contain information on safety and security perceptions.} to school made by children under the age of 12. The dataset comprises 2,999 observations, covering four travel modes: car, public transport, cycling, and walking. Walking is the most frequent mode, accounting for the majority of journeys (2,199 observations), whereas cycling is the least common (259 observations). In the following, we present the means of our variables presented in Table \ref{tab:descriptive_stats}, grouped into five categories of factors influencing children’s travel mode choice for school journeys (in Figure \ref{Fig:Corr} in Appendix \ref{Appendix_AdditionalFigure} we present a correlation matrix of selected variables).

\begin{table}[h]
	\centering
	\small
	\caption{Descriptive statistics by travel mode (average values per mode)}
	\begin{tabular}{lccccc}
		\hline
		 & Car & Public Transport & Cycling & Walking & Overall \\ \hline
		\multicolumn{6}{l}{\textit{Distance to school and travel time}}\\
		Travel time (minutes) & 11.70 & 24.00 & 10.74 & 11.15 & 14.39 \\
		Distance (km) & 4.08 & 3.94 & 1.53 & 0.75 & 2.57 \\
		Same municipality (yes in \%) & 73 & 75 & 92 & 99 & 85 \\
		\multicolumn{6}{l}{\textit{Built environment}}\\
		Urban (yes in \%) & 65 & 43 & 65 & 69 & 60 \\
		Suburban (yes in \%)& 20&21&20&18&20\\
		Rural (yes in \%) & 15 & 35 & 15 & 13 & 20 \\
		PT service level very good (yes in \%) & 9 & 5 & 9 & 15 & 10 \\
		PT service level good (yes in \%) & 11 & 7 & 9 & 18 & 11 \\
		PT service level moderate (yes in \%) & 25 & 15 & 29 & 28 & 24 \\
		PT service level poor (yes in \%) & 29 & 38 & 28 & 26 & 30 \\
		No PT service level (yes in \%) &25&34&25&13&24\\
		\multicolumn{6}{l}{\textit{Socio-demographic characteristics}}\\
		Age (years) & 8.86 & 9.08 & 9.81 & 8.88 & 9.16 \\
		Female (yes in \%) & 52 & 41 & 55 & 50 & 50 \\
		Nationality Swiss (yes in \%) & 69 & 72 & 76 & 71 & 72 \\
		French speaking (yes in \%) & 52 & 64 & 11 & 28 & 39 \\
		Italian speaking (yes in \%) & 10 & 5 & 3 & 2 & 5 \\
		German speaking (yes in \%) & 39&   31&   86&    69& 56\\
		\multicolumn{6}{l}{\textit{Parental and household characteristics}}\\
		Household size (number) & 4.15 & 4.16 & 4.36 & 4.29 & 4.24 \\
		Cars (number) & 1.67 & 1.69 & 1.53 & 1.44 & 1.58 \\
		Bikes  (number) & 3.47 & 3.80 & 4.35 & 3.86 & 3.87 \\
		Home ownership (yes in \%) & 58 & 67 & 63 & 51 & 59 \\
		\multicolumn{6}{l}{\textit{Weather}}\\
		Rain at reference date (yes in \%) & 24 & 19 & 16 & 22 & 20 \\
		\hline
		Observations & 280 & 261 & 259 & 2199 & 2999 \\\hline
	\end{tabular}
		\begin{tablenotes}[flushleft]
		\footnotesize
		\setstretch{1.2}
		\item \textit{Notes:} 'PT' denotes public transport.
	\end{tablenotes}
	\label{tab:descriptive_stats}
\end{table}

The average travel duration for school journeys is highest for public transport (24.00 minutes) and lowest for cycling (10.74 minutes).\footnote{Note that we ignore travel time for the analysis in Section \ref{Results} as mode choice for school journeys strongly influences the time traveled.} Noteworthy, travel distances for school journeys vary, with car journeys covering the longest distances on average (4.08 km), while, as expected, walking journeys cover the shortest distances (0.75 km).\footnote{Figure \ref{Fig:Hist} in \ref{Appendix_A} shows a histogram of distances by travel mode.} Moreover, we see that 85\% of all school journeys are made within the same municipality, with the highest share among the school journeys walked (99\%) and the lowest among car journeys (73\%). 

Related to built environment, we observe that, on average, 60\% of all children start their school journeys from urban areas, whereas 20\% start from rural areas---according to the definition of the Swiss Federal Office of Statistics, which is based on criteria related to density, size, and accessibility.\footnote{See also \href{https://www.bfs.admin.ch/bfs/de/home/statistiken/querschnittsthemen/raeumliche-analysen/raeumliche-gliederungen/raeumliche-typologien.html}{https://www.bfs.admin.ch/bfs/de/home/statistiken/querschnittsthemen/raeumliche-analysen/raeumliche-gliederungen/raeumliche-typologien.html}, accessed on March 24, 2025.} While most children who walk to school (69\%) come from urban areas, the share of public transport journeys to school originating from rural areas is relatively high (35\%), compared to the overall share of children from rural areas (20\%). Moreover, when the public transport service is classified as very good, according to the public transport service level indicator used to assess accessibility---as defined by the Federal Office for Spatial Development\footnote{See \href{https://www.are.admin.ch/are/de/home/medien-und-publikationen/publikationen/verkehr/ov-guteklassen-berechnungsmethodik-are.html}{https://www.are.admin.ch/are/de/home/medien-und-publikationen/publikationen/verkehr/ov-guteklassen-berechnungsmethodik-are.html}, accessed on March 24, 2025.}---the share of walked journeys is relatively high (15\%) compared to the overall share of this group (10\%). On the other hand, the share of journeys made by public transport to school is higher in areas with poor public transport service levels (38\%) than the overall share of this group (30\%).

Regarding socio-demographic characteristics, we see in Table \ref{tab:descriptive_stats} that no remarkable age differences exist in the averages between children walking to school (8.88) and those who were driven by car (8.86). The same occurs for gender, where the share of girls driven by car amounts to 52\% and is only slightly higher compared to the overall share of girls (50\%). Moreover, the proportion of Swiss children is the lowest (69\%) among children using the car to drive to and the highest (76\%) among those cycling to school. Moreover, the share of car journeys to school is higher among the French (52\%) and Italian (10\%) speaking regions compared to the overall shares of these regions (36\% and 5\%, respectively). On the other hand, the share of journeys walked (69\%) or cycled (86\%) to school is higher in the German speaking region compared to the overall share of this region (56\%).

Regarding parental and household characteristics, we mainly observe differences between the four classes in terms of bike and car ownership. The number of cars is higher (1.67 and 1.69, respectively) in households of children who travel to school by car or public transport, compared to the overall average of 1.58 cars per household in our sample. Conversely, the number of bikes is higher among children who cycle to school (4.35) compared to the overall average number of bikes in the sample (3.87).

Finally, regarding weather, the share of school journeys made by bicycle (16\%) is lower on rainy days compared to the overall share of rainy days (20\%).

\section{Results}\label{Results}

The following section presents the results of the travel mode choice prediction for school journeys. First, we report the model's performance in distinguishing between active (i.e., walking and cycling) and non-active (i.e., car and public transport) travel modes for school journeys. Second, we provide performance metrics and variable importance for the prediction of car usage vs other travel modes for school journeys. Additional analyses focusing on the prediction of i) the four travel mode classes walking, cycling, car, and public transport, ii) children who walk to school, and iii) car vs. public transport are presented in Appendix \ref{Appendix_A}, Appendix \ref{Appendix_B}, and Appendix \ref{Appendix_C}, respectively.

\subsection{Active versus non-active travel}

\begin{table}[H]
	\centering
	\caption{Overall and class-specific prediction rates with 95\% prediction intervals (active versus non-active travel)}
	\begin{tabular}{lccc}
		\toprule
		Mode & Mean Accuracy & 95\% PI Lower & 95\% PI Upper \\
		\midrule
		\textbf{Overall} & \textbf{0.8727} & \textbf{0.8682} & \textbf{0.8771} \\
		Active modes & 0.874 & 0.867 & 0.881 \\
		Non-active modes & 0.871 & 0.865 & 0.878 \\
		\hline
	\end{tabular}
			\begin{tablenotes}[flushleft]
		\footnotesize
		\setstretch{1.2}
		\item \textit{Notes:} 'PI' denotes prediction interval.
	\end{tablenotes}
	\label{tab:prediction_rates_active}
\end{table}

In Table \ref{tab:prediction_rates_active} we see that the random forest model predicting travel mode choice for school journeys achieved an overall classification accuracy of 87.27\%, with a 95\% prediction interval of [86.82\%; 87.71\%], indicating a high level of predictive performance. When broken down by mode category, the model classifies active modes with an accuracy of 87.4\% (95\%-prediction interval: [86.7\%; 88.1\%]) and non-active modes with an accuracy of 87.1\% (95\%-prediction interval: [86.5\%; 87.8\%]). Our results suggest that the model performs consistently well across both categories, with only marginal differences in classification accuracy between active and non-active travel modes.

\begin{figure}[H] 
	\includegraphics[scale=1]{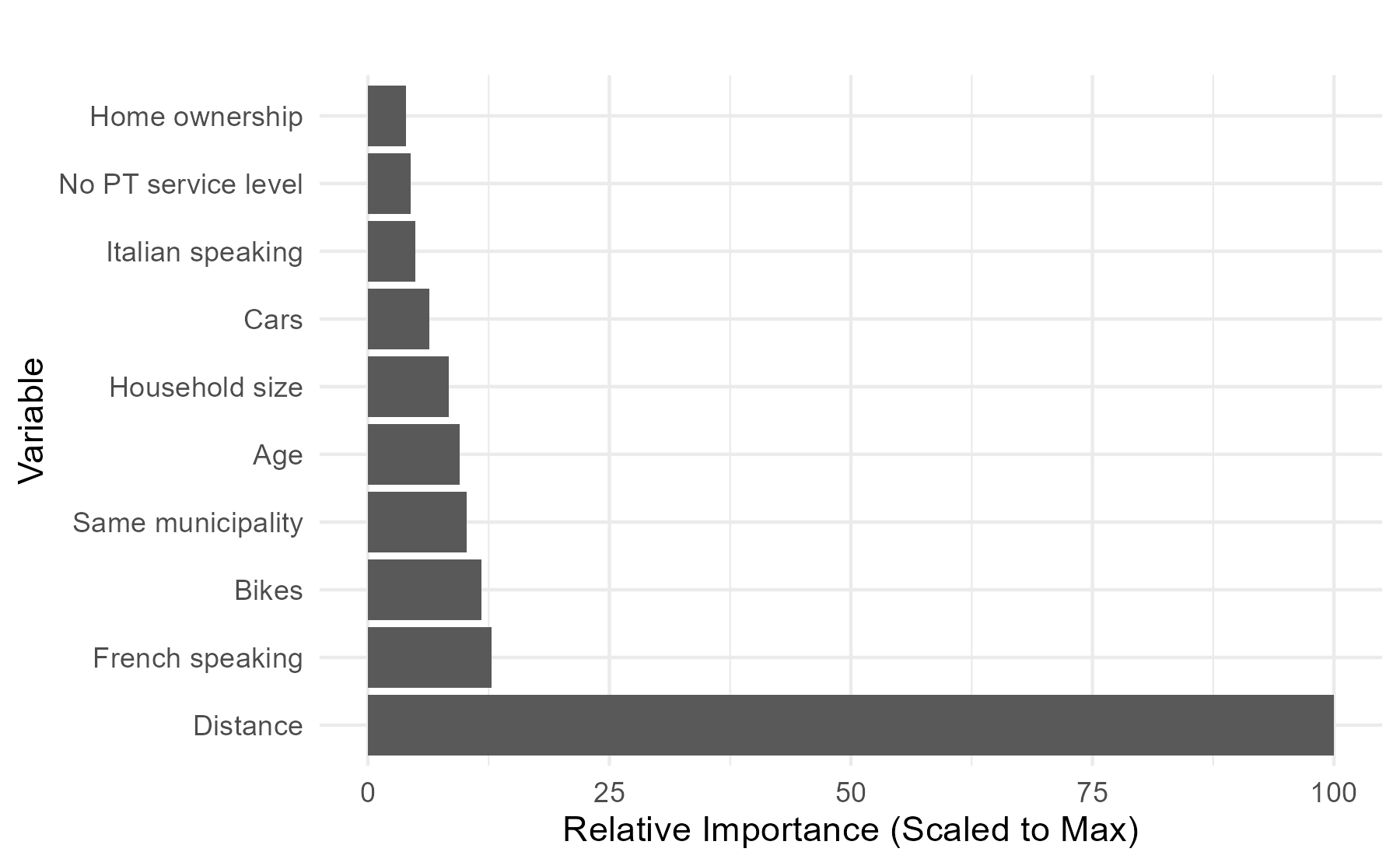}
	\centering \caption{Variable importance plot for the prediction of the children travel mode choice, comparing active and non-active travel modes. We compute the variable importance using the mean decrease in the Gini index and express it relative to the maximum.} \label{Fig:VariableImportance_Active}
\end{figure}

Distance is the most important predictor, with an average Mean Decrease Gini (MDG) of 151.0, serving as the baseline (100\%) for relative importance in Figure \ref{Fig:VariableImportance_Active}. Other influential variables include French as main language (MDG = 19.4, 12.8\% relative importance), number of bikes (MDG = 17.7, 11.8\%), same municipality (MDG = 15.5, 10.3\%), age (MDG = 14.4, 9.5\%), and household size (MDG = 12.6, 8.4\%).

\begin{figure}[H] 
	\includegraphics[scale=1]{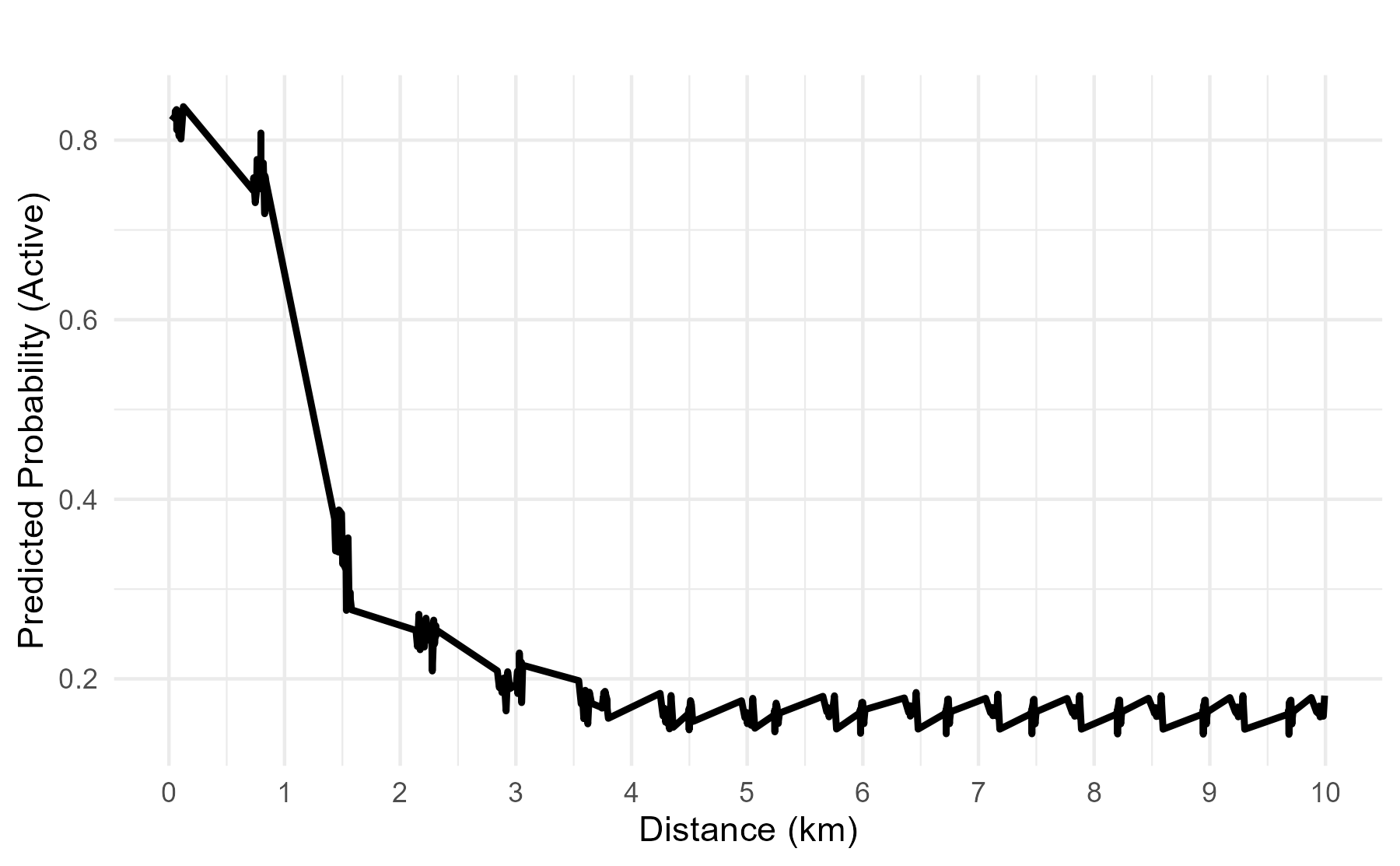}
	\centering \caption{Average effect of distance (<10 km)} \label{Fig:Dist_Active}
\end{figure}

In Figure \ref{Fig:Dist_Active}, we show the partial dependence plot for distance. Partial dependence plots illustrate the marginal effect of a predictor variable on the predicted outcome, averaging over the influence of all other variables in the model. We observe that across all 100 repetitions, as distance increases, the predicted probability of choosing an active mode for school journeys decreases. Moreover, Figure \ref{Fig:Dist_Active} indicates that the probability of walking or biking to school falls below 0.5 between 1 km and 1.5 km distance from home to school.

\subsection{Predicting the use of parental cars for school journeys}

Table \ref{tab:prediction_rates_auto} further reports the model's predictive performance when distinguishing between car use and all other school travel modes for school journeys. The overall classification accuracy in this binary setup is 78.97\%, with a 95\% prediction interval of [78.28\%; 79.67\%]. The model classifies car journeys with an accuracy of 81.8\% (95\%-prediction interval: [80.7\%; 82.9\%]) and other modes with an accuracy of 76.3\% (95\%-prediction interval: [75.2\%; 77.5\%]). These results suggest that the model performs slightly better at identifying car journeys compared to other travel modes for school journeys, although predictive performance remains relatively balanced across the two categories.

\begin{table}[H]
	\centering
	\caption{Overall and class-specific prediction rates with 95\% prediction intervals (car versus other modes)} 
	\begin{tabular}{lccc}
		\toprule
		Mode & Mean Accuracy & 95\% PI Lower & 95\% PI Upper \\
		\midrule
		\textbf{Overall} & \textbf{0.7897} & \textbf{0.7828} & \textbf{0.7967} \\
		Car & 0.818 & 0.807 & 0.829 \\  
		Other modes & 0.763 & 0.752 & 0.775 \\ 
		\hline
	\end{tabular}
	\begin{tablenotes}[flushleft]
		\footnotesize
		\setstretch{1.2}
		\item \textit{Notes:} 'PI' denotes prediction interval.
	\end{tablenotes}
	\label{tab:prediction_rates_auto}
\end{table}

Distance remains the most important predictor, with an average Mean Decrease Gini (MDG) of 57.9, serving as the baseline (100\%) for relative importance in Figure \ref{Fig:VariableImportance}.\footnote{Note that the proportion of children using public transport for school journeys is relatively small compared to the other modes group.} Other influential variables include number of bikes (MDG = 13.0, 22.5\% relative importance) and age (MDG = 10.5, 18.1\%).

\begin{figure}[H] 
	\includegraphics[scale=1]{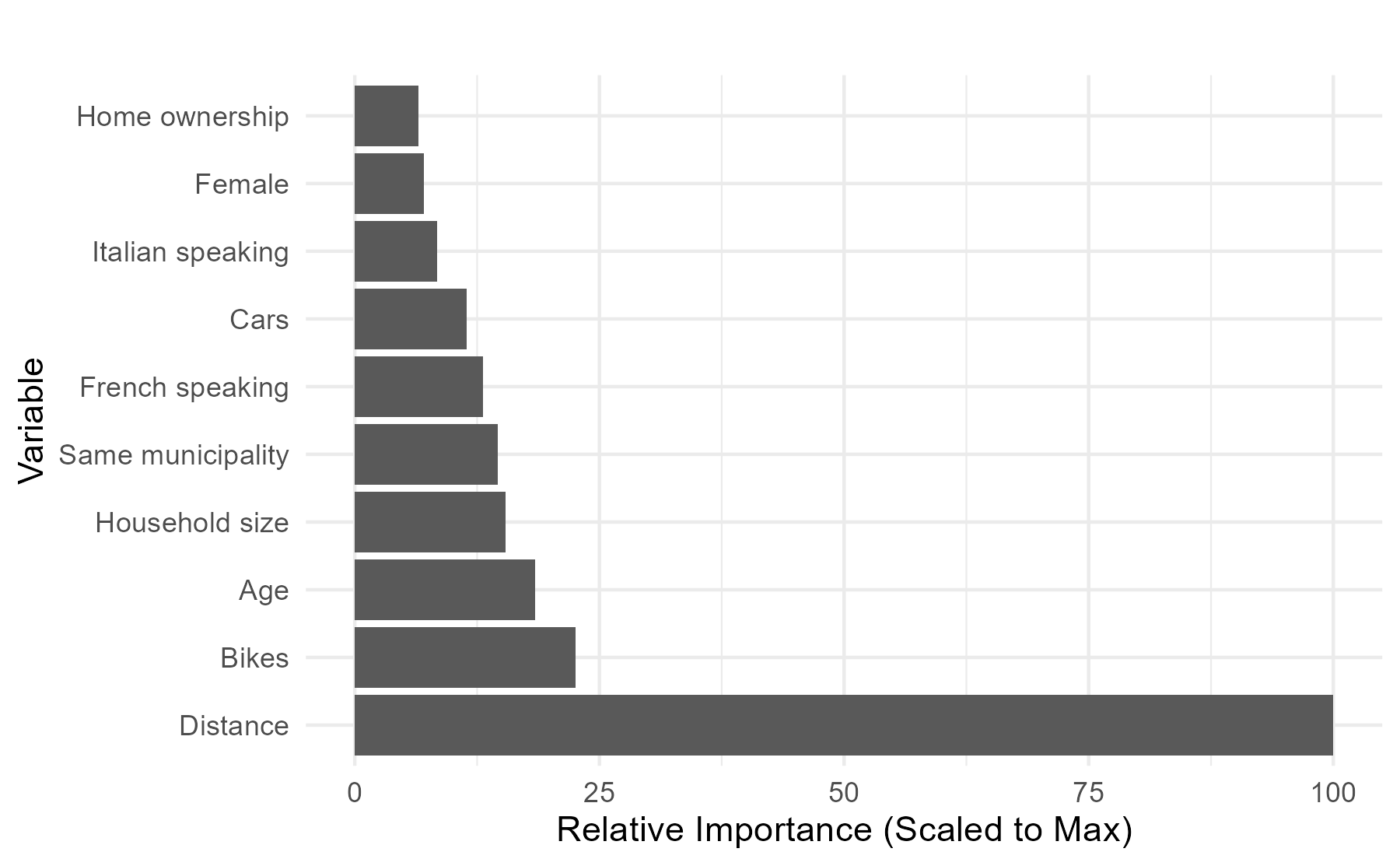}
	\centering \caption{Variable importance plot for the prediction of the children's travel mode choice for school journeys, comparing children who drive and those who do not drive to school. We compute the variable importance using the mean decrease in the Gini index and express it relative to the maximum.} \label{Fig:VariableImportance_Car}
\end{figure}

In Figure \ref{Fig:Dist_Active}, we again show the partial dependence plot for distance. We observe that across all 100 repetitions, as distance increases, the predicted probability of choosing car usage increases. The probability of no car usage increases steadily up to a distance of 2 km from home to school.

\begin{figure}[H] 
	\includegraphics[scale=1]{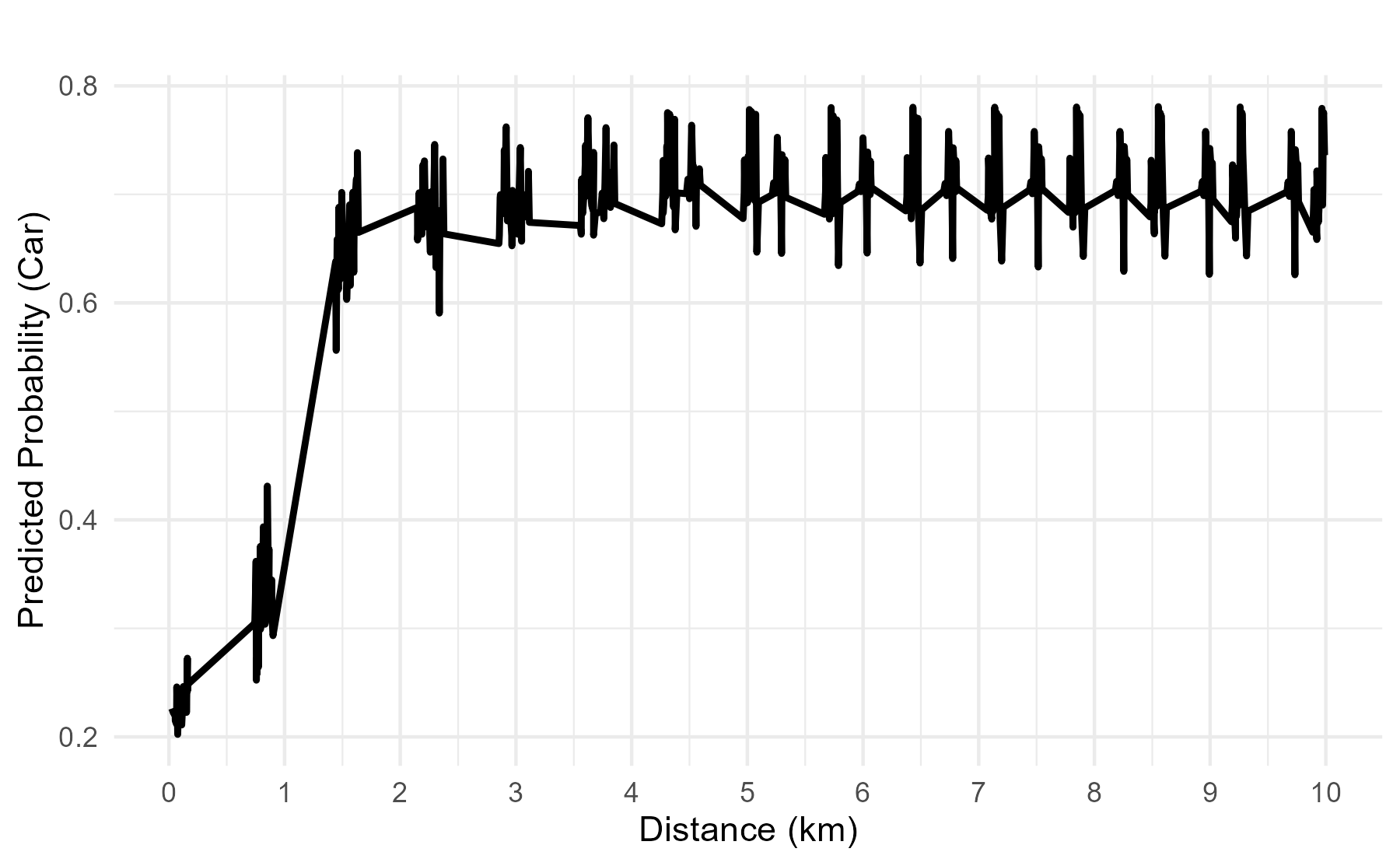}
	\centering \caption{Average effect of distance (<10 km)} \label{Fig:Dist_Car}
\end{figure}

\subsection{Predicting the use of parental cars for school journeys when excluding distance}

To better understand the influence of factors other than distance on car usage, we re-estimated the random forest model excluding distance as a predictor. Table \ref{tab:prediction_rates_auto} presents the model’s predictive performance when distinguishing between car use and all other school travel modes for school journeys under this new specification. The overall classification accuracy decreases to 70.80\%, with a 95\% prediction interval of [69.94\%; 71.67\%]. The model classifies car journeys with an accuracy of 71.4\% (95\%-prediction interval: [70.2\%; 72.7\%]) and other modes with an accuracy of 70.2\% (95\%-prediction interval: [68.9\%; 71.6\%]). Although performance drops compared to the full model, the results indicate that reasonable classification is still achievable based solely on non-distance-related variables.

\begin{table}[H]
	\centering
	\caption{Overall and class-specific prediction rates with 95\% prediction intervals (car versus other modes, excluding distance)} 
	\begin{tabular}{lccc}
		\toprule
		Mode & Mean Accuracy & 95\% PI Lower & 95\% PI Upper \\
		\midrule
		\textbf{Overall} & \textbf{0.7080} & \textbf{0.6994} & \textbf{0.7167} \\
		Car & 0.714 & 0.702 & 0.727 \\  
		Other modes & 0.702 & 0.689 & 0.716 \\ 
		\hline
	\end{tabular}
	\begin{tablenotes}[flushleft]
		\footnotesize
		\setstretch{1.2}
		\item \textit{Notes:} 'PI' denotes prediction interval. Results are based on a model excluding distance as a predictor.
	\end{tablenotes}
	\label{tab:prediction_rates_auto_ohneDist}
\end{table}

\begin{figure}[H] 
	\includegraphics[scale=1]{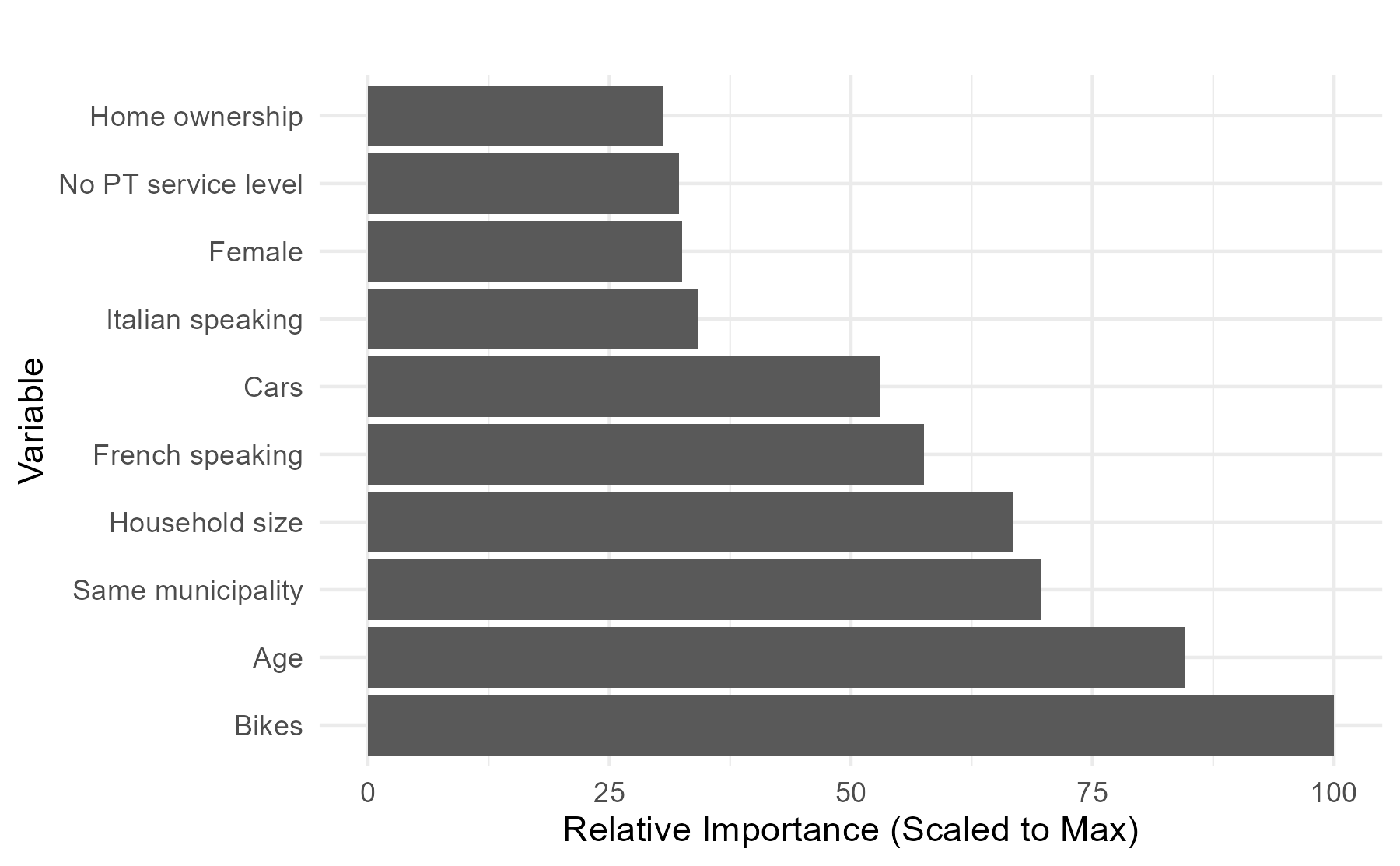}
	\centering \caption{Variable importance plot for the prediction of  children's travel mode choice for school journeys, comparing children who drive and those who do not drive to school, excluding distance as a predictor. We compute the variable importance using the mean decrease in the Gini index and express it relative to the maximum.} \label{Fig:VariableImportance_Car_ohneDist}
\end{figure}

When excluding distance from the model, the number of bikes becomes the most important predictor of children's travel mode choice for school journeys, with an average Mean Decrease Gini (MDG) of 18.8, serving as the baseline (100\%) for relative importance in Figure \ref{Fig:VariableImportance_Car_ohneDist}. Other influential variables include age (MDG = 15.9, 84.5\% relative importance), living in the same municipality as the school (MDG = 13.1, 69.7\%), household size (MDG = 12.6, 66.9\%), and speaking French as the main household language (MDG = 10.8, 57.6\%).

\section{Discussion and Conclusion}\label{Discussion}

This study assessed the predictive accuracy of children's mode choice for school journeys, with a focus on active travel and car use. We found that active travel modes (walking and biking) were predicted with high accuracy (87.27\%), while the model achieved a slightly lower accuracy of 78.97\% when distinguishing between car use and other modes. Even when distance was excluded from the model---a key factor in mode choice decisions for school journeys---the model still performed reasonably well, with an overall accuracy of 70.80\%. Across all specifications, distance to school consistently emerged as by far the most influential predictor. Other important factors included the child’s age, the number of bicycles in the household, language region, household size, and whether the school was located in the same municipality.

Our study’s findings closely align with current Swiss efforts to reduce parental cars for school journeys. Relevant initiatives include local campaigns in Kölliken (AG)\footnote{See \href{https://www.luzernerzeitung.ch/aargau/wyna-suhre/koelliken-starke-kinder-gehen-zu-fuss-koelliken-sagt-den-elterntaxis-den-kampf-an-ld.2752194}{https://www.luzernerzeitung.ch/aargau/wyna-suhre/koelliken-starke-kinder-gehen-zu-fuss-koelliken-sagt-den-elterntaxis-den-kampf-an-ld.2752194}, accessed on April 1, 2025.} and Walchwil (ZG)\footnote{See \href{https://www.zentralplus.ch/regionales-leben/diese-zuger-gemeinde-geht-erfolgreich-gegen-elterntaxis-vor-2703355/}{https://www.zentralplus.ch/regionales-leben/diese-zuger-gemeinde-geht-erfolgreich-gegen-elterntaxis-vor-2703355/}, accessed on April 1, 2025.}, as well as public transport voucher schemes in Lucerne \citep[see][]{balthasar2024fare}. Furthermore, the key factors identified in this study---such as school distance, child’s age, and household characteristics---are directly relevant to national programs like Walk to School\footnote{See \href{https://www.verkehrsclub.ch/schulweg/walktoschool}{https://www.verkehrsclub.ch/schulweg/walktoschool}, accessed on April 1, 2025.} (targeting teachers) and Pedibus\footnote{See \href{https://www.verkehrsclub.ch/schwerpunkte/ein-tausendfuessler-jubiliert}{https://www.verkehrsclub.ch/schwerpunkte/ein-tausendfuessler-jubiliert}, accessed on April 1, 2025.} (targeting parents), both supported by the Swiss Association for Transport and Environment.

However, our results do not establish a causal relationship---for example, between the policy measures discussed above and children's mode choice for school journeys. Future research should therefore aim to examine the causal effects of such measures on children’s travel behavior. In this regard, our study serves as a targeted source of inspiration for identifying potential confounding variables.

A limitation of our study is that the Swiss Mobility and Transport Microcensus (MTMC) does not collect data on parents’ perceived safety and risk factors. Investigating these perceptions within the Swiss context represents an important topic for future research. Another important direction for future research is the distinction between escorted and independent school journeys.

	\newpage
	\bigskip
	
	\bibliographystyle{sageh}
	\bibliography{SchoolTravelModeChoice.bib}
	
	\newpage
	\bigskip

\begin{appendix}
		
		\numberwithin{equation}{section}
		\counterwithin{figure}{section}
		\noindent \textbf{\LARGE Appendices}

\section{Additional Figures}\label{Appendix_AdditionalFigure}

\begin{figure}[H] 
	\includegraphics[scale=1]{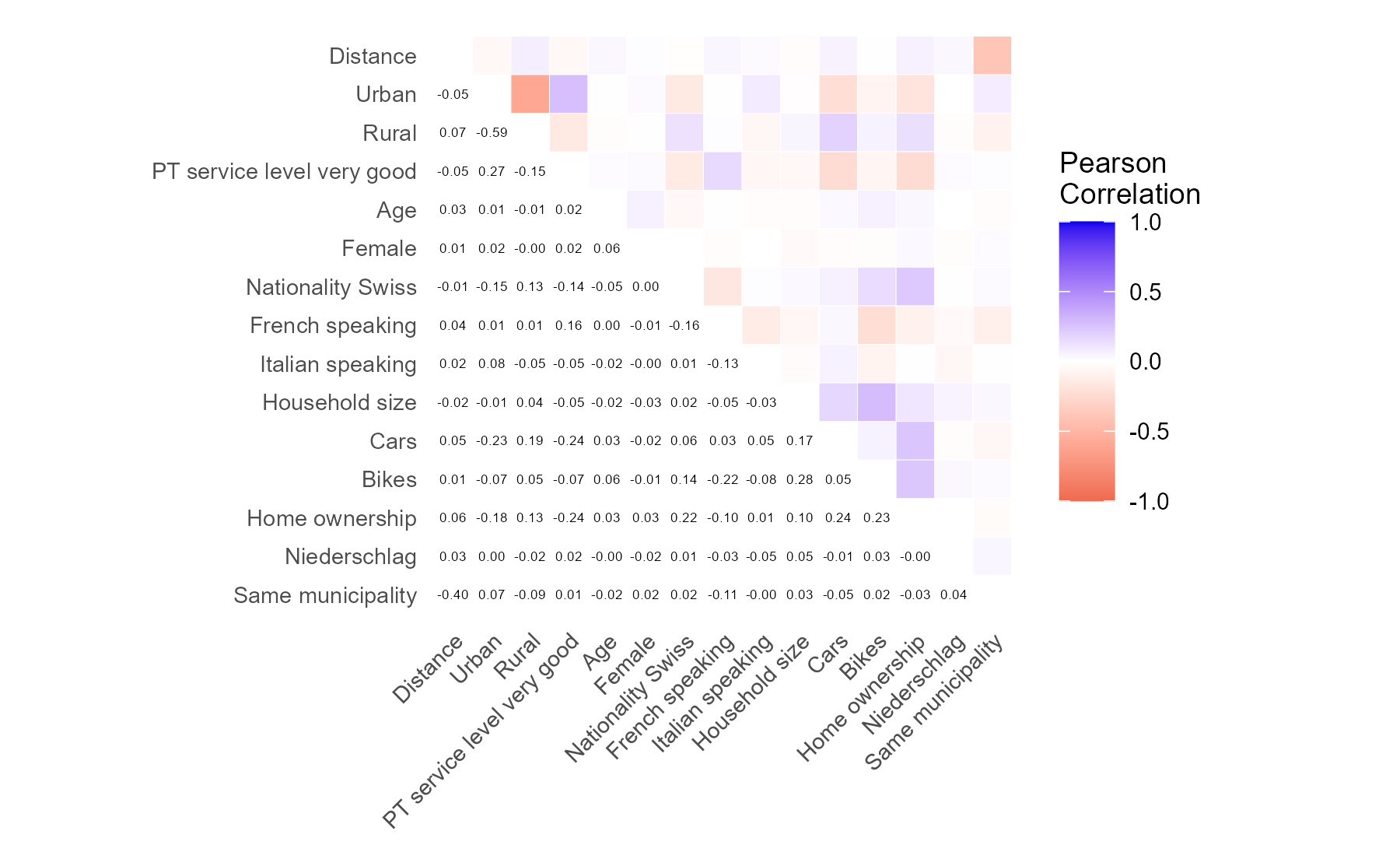}
	\centering \caption{Correlation matrix of selected predictors.} \label{Fig:Corr}
\end{figure}

\begin{figure}[H] 
	\includegraphics[scale=1]{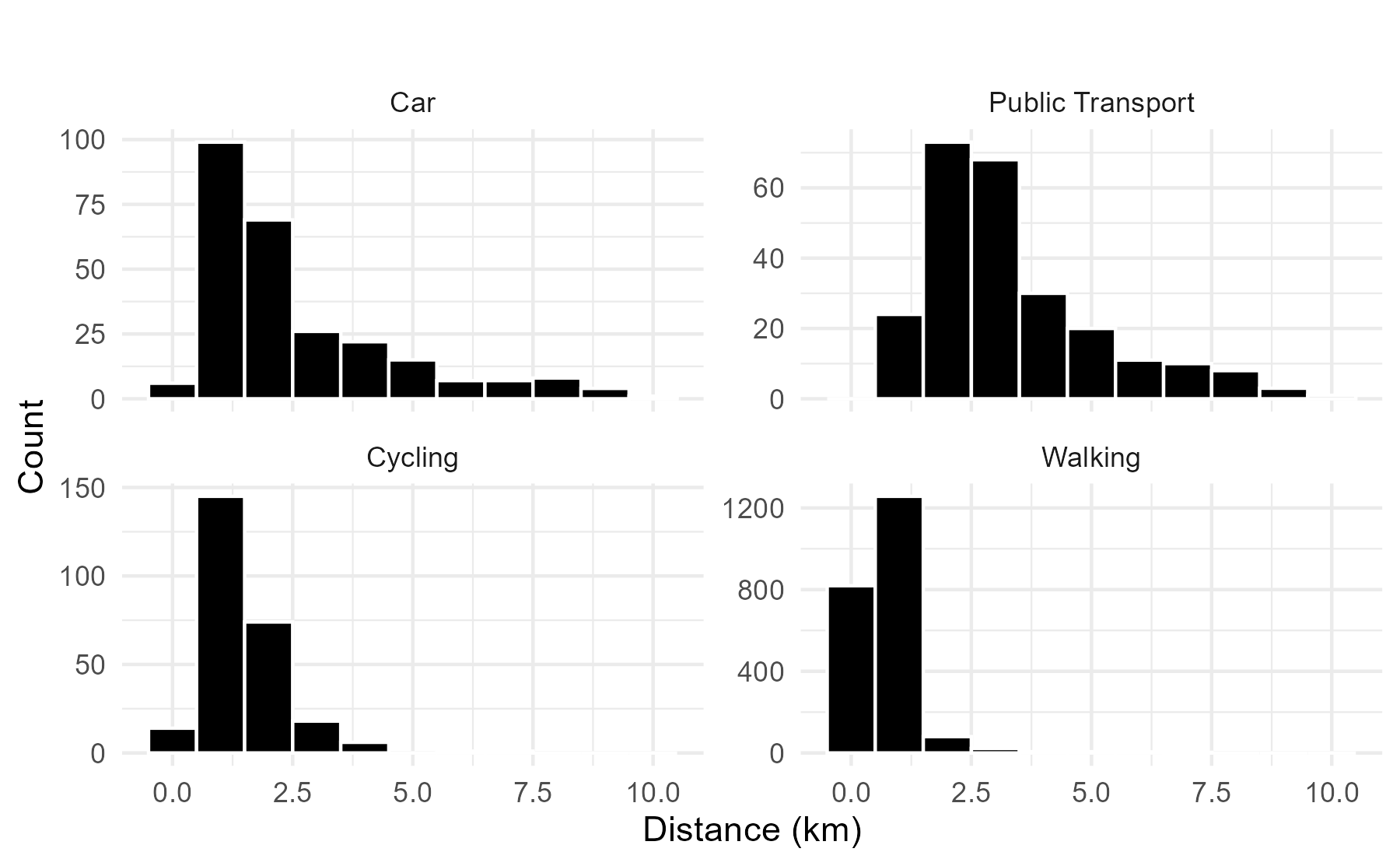}
	\centering \caption{Histogram of Distances (<10km) by Travel Mode} \label{Fig:Hist}
\end{figure}

\section{Predicting travel mode choice}\label{Appendix_A}

\begin{table}[H]
	\centering
	\caption{Overall and Mode-Specific Prediction Rates with 95\% Prediction Intervals (walking, cycling, public transport, and car)}
	\begin{tabular}{lccc}
		\toprule
		Mode & Mean Accuracy & 95\% PI Lower & 95\% PI Upper \\
		\midrule
		\textbf{Overall} & \textbf{0.663} & \textbf{0.6571} & \textbf{0.6689} \\
		Walking & 0.764 & 0.751 & 0.776 \\
		Cycling & 0.703 & 0.690 & 0.716 \\
		Public Transport & 0.677 & 0.663 & 0.690 \\
		Car & 0.513 & 0.499 & 0.528 \\
		\hline
	\end{tabular}
	\begin{tablenotes}[flushleft]
		\footnotesize
		\setstretch{1.2}
		\item \textit{Notes:} 'PI' denotes prediction interval.
	\end{tablenotes}
	\label{tab:prediction_rates_all}
\end{table}

As we can see in Table \ref{tab:prediction_rates_all}, the random forest model when predicting the four classes of travel mode choice (i.e., walking, cycling, public transport, and car) achieved an overall classification accuracy of 66.56\% with a 95\%-prediction interval of [66.01\%; 67.11\%], indicating a moderate level of predictive performance in distinguishing between travel modes. Mode-specific accuracy varied considerably, with the highest classification rate observed for walking (76.3\%, 95\%-prediction interval: [75.1\%; 77.6\%]), followed by cycling (70.4\%, 95\%-prediction interval: [69.1\%; 71.7\%]) and public transport (67.3\%, 95\%-prediction interval: [66.0\%; 68.7\%]). The lowest accuracy was found for car travel at 52.5\% (95\%-prediction interval: [51.0\%; 53.9\%]), suggesting greater classification difficulty for this mode. These results highlight that walking journeys are more easily distinguishable based on the selected predictors, whereas motorized travel modes, particularly car journeys, exhibit greater classification uncertainty.

\begin{figure}[H] 
	\includegraphics[scale=1]{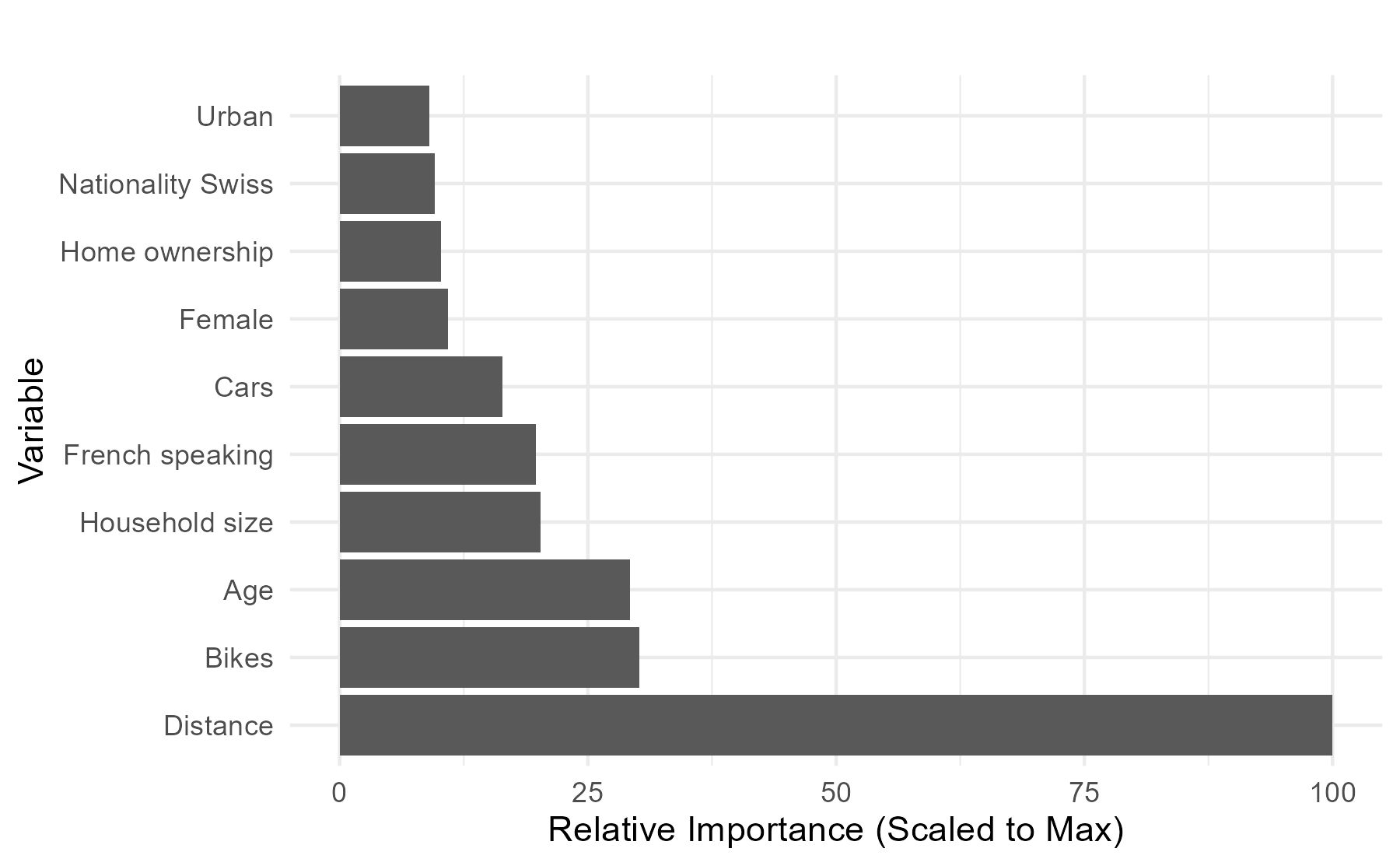}
	\centering \caption{Variable importance plot for the prediction of the children travel mode choice. We compute the variable importance using the mean decrease in the Gini index and express it relative to the maximum.} \label{Fig:VariableImportance}
\end{figure}

Distance remains the most important predictor, with an average Mean Decrease Gini (MDG) of 134.0, serving as the baseline (100\%) for relative importance in Figure \ref{Fig:VariableImportance}. Other influential variables include number of bikes (MDG = 40.5, 30.2\% relative importance), and age (MDG = 39.2, 29.2\%). 

\section{Predicting children walking to school}\label{Appendix_B}

\begin{table}[H]
	\centering
	\caption{Overall and class-specific prediction rates with 95\% prediction intervals (walking versus non-walking children)}
	\begin{tabular}{lccc}
		\hline
		Mode & Mean Accuracy & 95\% PI Lower & 95\% PI Upper \\
		\midrule
		\textbf{Overall} & \textbf{0.8666} & \textbf{0.8630} & \textbf{0.8703} \\
		Walking & 0.877 & 0.871 & 0.883 \\ 
		Other modes & 0.857 & 0.851 & 0.862 \\ 
		\hline
	\end{tabular}
	\begin{tablenotes}[flushleft]
		\footnotesize
		\setstretch{1.2}
		\item \textit{Notes:} 'PI' denotes prediction interval.
	\end{tablenotes}
	\label{tab:prediction_rates_fuss}
\end{table}

\begin{figure}[H] 
	\includegraphics[scale=1]{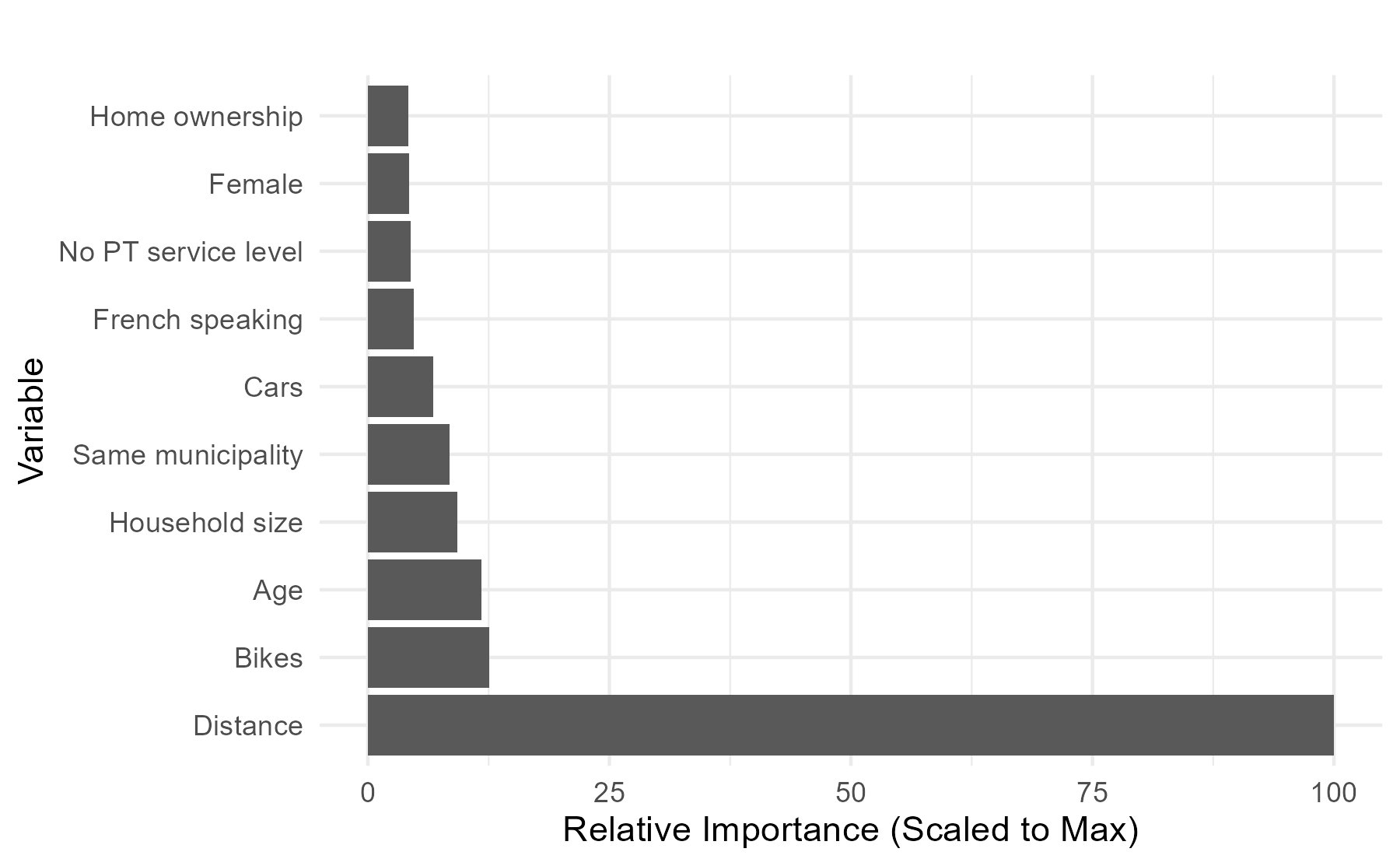}
	\centering \caption{Variable importance plot for the prediction of the children's travel mode choice, comparing children who walk and those who do not walk. We compute the variable importance using the mean decrease in the Gini index and express it relative to the maximum.} \label{Fig:VariableImportance_Walk}
\end{figure}

\section{Predicting car vs. public transport}\label{Appendix_C}

\begin{table}[H]
	\centering
	\caption{Overall and class-specific prediction rates with 95\% prediction intervals (public transport versus car usage)}
	\begin{tabular}{lccc}
		\hline
		Mode & Mean Accuracy & 95\% PI Lower & 95\% PI Upper \\
		\midrule
		\textbf{Overall} & \textbf{0.7230} & \textbf{0.7148} & \textbf{0.7312} \\
		Car & 0.741 & 0.728 & 0.755 \\
		Public Transport & 0.708 & 0.695 & 0.721 \\
		\hline
	\end{tabular}
	\begin{tablenotes}[flushleft]
		\footnotesize
		\setstretch{1.2}
		\item \textit{Notes:} 'PI' denotes prediction interval.
	\end{tablenotes}
	\label{tab:prediction_rates_car_vs_pt}
\end{table}

\begin{figure}[H] 
	\includegraphics[scale=1]{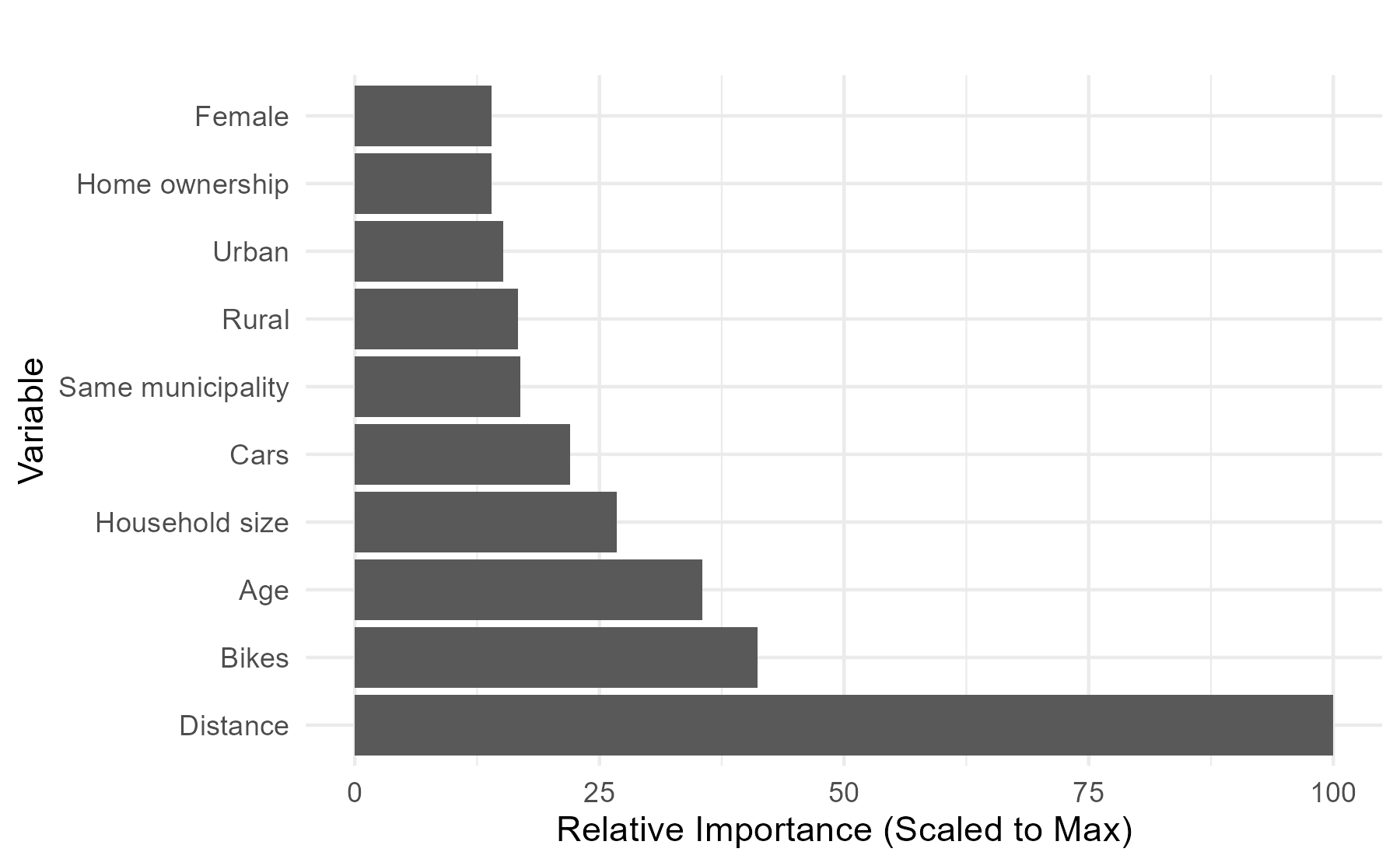}
	\centering \caption{Variable importance plot for the prediction of the children's travel mode choice, comparing children who use (only) car or public transport to travel to school. We compute the variable importance using the mean decrease in the Gini index and express it relative to the maximum.} \label{Fig:VariableImportance_car_pt}
\end{figure}

	\end{appendix}
\end{document}